# The Gaussian MAC with Conferencing Encoders

Shraga I. Bross
Bar Ilan University
Ramat Gan, 52900, Israel
brosss@macs.biu.ac.il

Amos Lapidoth
ETH Zurich
8092 Zurich, Switzerland
lapidoth@isi.ee.ethz.ch

Michèle A. Wigger
ETH Zurich
8092 Zurich, Switzerland
wigger@isi.ee.ethz.ch

*Abstract*—We derive the capacity region of the Gaussian version of Willems's two-user MAC with conferencing encoders. This setting differs from the classical MAC in that, prior to each transmission block, the two transmitters can communicate with each other over noise-free bit-pipes of given capacities.

The derivation requires a new technique for proving the optimality of Gaussian input distributions in certain mutual information maximizations under a Markov constraint.

We also consider a Costa-type extension of the Gaussian MAC with conferencing encoders. In this extension, the channel can be described as a two-user MAC with Gaussian noise and Gaussian interference where the interference is known non-causally to the encoders but not to the decoder. We show that as in Costa's setting the interference sequence can be perfectly canceled, i.e., that the capacity region without interference can be achieved.

## I. INTRODUCTION

We consider a communication scenario known as the MAC with conferencing encoders where two transmitters wish to transmit independent messages to a single receiver. Prior to each transmission block, the two encoders are allowed to hold a *conference*, i.e., they can communicate with each other over noise-free bit-pipes of given capacities. Special cases are the classical multiple-access setting, where the encoders are ignorant of each others messages (both bit-pipes of zero capacities); the fully-cooperative setting (both bit-pipes of infinite capacities); and the asymmetric message sets setting, where one of the encoders is fully cognizant of the message the other encoder intends to send (the pipe from the cognizant transmitter to the non-cognizant transmitter of zero capacity and the other pipe of infinite capacity).

The MAC with conferencing encoders was introduced by Willems in [1], who also derived the capacity region for the discrete memoryless setting. Here we derive the capacity region for the Gaussian setting under average power constraints. The achievability part is very similar to the one in [1]. The converse, however, requires a novel tool first derived in [4] for proving that Gaussian distributions maximize certain mutual information expressions under a Markovity-constraint. For such maximization problems the traditional approach of proving the optimality of Gaussian distributions by employing the Max-Entropy Theorem [3, Theorem 12.1.1.] or a conditional version thereof [5] fails. The reason is that replacing a non-Gaussian vector satisfying the Markovity condition by a Gaussian vector of the same covariance matrix may result in a Gaussian vector that violates the Markovity condition.

We also consider an additional scenario where the received signal is also corrupted by an additive Gaussian interference sequence which is non-causally known to both encoders but not to the decoder. We show that, even though the decoder is non-informed, the interference can be perfectly canceled in the sense that the capacity region of the setting without interference is achievable also in this setting with interference.

We next describe the channel models more precisely and then proceed to state our results.

The goal of the transmission is that Transmitters 1 and 2 convey their messages $M_1$ and $M_2$ to the receiver. The messages $M_1$ and $M_2$ are assumed to be independent and uniformly distributed over the sets $\mathcal{M}_1 = \{1, \ldots, \lfloor e^{nR_1} \rfloor\}$ and $\mathcal{M}_2 = \{1, \ldots, \lfloor e^{nR_2} \rfloor\}$. Here $n$ denotes the block-length, and $R_1$ and $R_2$ denote the rates of transmission in nats per channel use.

Prior to each block of $n$ channel uses, the two encoders hold a conference, i.e., they exchange information over $k$ uses of the pipes. The pipes are assumed to be

- perfect in the sense that any input symbol to a pipe is available immediately and error-free at the output of the pipe; and
- of limited throughputs $C_{12}$ and $C_{21}$, in the sense that when the $k$ inputs to the pipe from Transmitter 1 to Transmitter 2 take values in the sets $\mathcal{V}_{1,1}, \ldots, \mathcal{V}_{1,k}$ and the $k$ inputs to the pipe from Transmitter 2 to Transmitter 1 take values in the sets $\mathcal{V}_{2,1}, \ldots, \mathcal{V}_{2,k}$, then

$$\sum_{\ell=1}^{k} \log |\mathcal{V}_{1,\ell}| \leq nC_{12} \quad \text{and} \quad \sum_{\ell=1}^{k} \log |\mathcal{V}_{2,\ell}| \leq nC_{21}. \quad (1)$$

Here and throughout all logarithms are natural logarithms.

Note that the communication over the pipes is assumed to be held in a conferencing way, so that the $\ell$-th inputs $V_{1,\ell} \in \mathcal{V}_{1,\ell}$ and $V_{2,\ell} \in \mathcal{V}_{2,\ell}$ can depend on the respective messages as well as on the past observed pipe-outputs:

$$V_{1,\ell} = f_{1,\ell}(M_1, V_{2,1}, \ldots, V_{2,\ell-1}), \quad (2)$$
$$V_{2,\ell} = f_{2,\ell}(M_2, V_{1,1}, \ldots, V_{1,\ell-1}), \quad (3)$$

for some given sequences of encoding functions $\{f_{1,\ell}\}_{\ell=1}^{k}$ and $\{f_{2,\ell}\}_{\ell=1}^{k}$ where

$$f_{1,\ell}: \quad \mathcal{M}_1 \times \mathcal{V}_{2,1} \times \ldots \times \mathcal{V}_{2,\ell-1} \longrightarrow \mathcal{V}_{1,\ell}, \quad (4)$$
$$f_{2,\ell}: \quad \mathcal{M}_2 \times \mathcal{V}_{1,1} \times \ldots \times \mathcal{V}_{1,\ell-1} \longrightarrow \mathcal{V}_{2,\ell}. \quad (5)$$

The research of Michèle A. Wigger was supported by the Swiss National Science Foundation under Grant 200021-111863/1.

We define an $(n, C_{12}, C_{21})$-*conference* to be the collection of an integer number $k$, two sets of input alphabets $\{\mathcal{V}_{1,1}, \ldots, \mathcal{V}_{1,k}\}$ and $\{\mathcal{V}_{2,1}, \ldots, \mathcal{V}_{2,k}\}$, and two sets of encoding functions $\{f_{1,1}, \ldots, f_{1,k}\}$ and $\{f_{2,1}, \ldots, f_{2,k}\}$ as in (4) and (5), where $n, C_{12}, C_{21}, k$, and the sets $\{\mathcal{V}_{1,1}, \ldots, \mathcal{V}_{1,k}\}$ and $\{\mathcal{V}_{2,1}, \ldots, \mathcal{V}_{2,k}\}$ satisfy (1).

After the conference, Transmitter 1 is cognizant of the sequence $\mathbf{V}_2 = (V_{2,1}, \ldots, V_{2,k})$ and Transmitter 2 is cognizant of the sequence $\mathbf{V}_1 = (V_{1,1}, \ldots, V_{1,k})$. The channel input sequences $\mathbf{X}_1 = (X_{1,1}, \ldots, X_{1,n})$ and $\mathbf{X}_2 = (X_{2,1}, \ldots, X_{2,n})$ can then be described with encoding functions $\varphi_1^{(n)}$ and $\varphi_2^{(n)}$ as

$$\mathbf{X}_1 = \varphi_1^{(n)}(M_1, \mathbf{V}_2), \tag{6}$$
$$\mathbf{X}_2 = \varphi_2^{(n)}(M_2, \mathbf{V}_1), \tag{7}$$

where

$$\varphi_1^{(n)}: \mathcal{M}_1 \times \mathcal{V}_{2,1} \times \ldots \times \mathcal{V}_{2,k} \longrightarrow \mathbb{R}^n, \tag{8}$$
$$\varphi_2^{(n)}: \mathcal{M}_2 \times \mathcal{V}_{1,1} \times \ldots \times \mathcal{V}_{1,k} \longrightarrow \mathbb{R}^n. \tag{9}$$

Additionally, we impose an average block power constraint on both channel input sequences:

$$\frac{1}{n}\mathsf{E}\left[\sum_{t=1}^n (X_{\nu,t})^2\right] \leq P_\nu, \quad \nu \in \{1,2\}. \tag{10}$$

The multiple-access channel is described as follows. For given discrete-time $t$ and channel inputs $x_{1,t}, x_{2,t} \in \mathbb{R}$, the time $t$ channel output $Y_t$ is

$$Y_t = x_{1,t} + x_{2,t} + Z_t, \tag{11}$$

where $\{Z_t\}$ models the noise corrupting the channel and is given by a sequence of independent and identically distributed (IID) zero-mean Gaussian random variables of variance $\sigma^2 > 0$.

Based on the output sequence $\mathbf{Y} = (Y_1, \ldots, Y_n)$ the decoder applies a decoding function $\phi^{(n)}$,

$$\phi^{(n)}: \mathbb{R}^n \to \mathcal{M}_1 \times \mathcal{M}_2, \tag{12}$$

to produce the message estimates $\hat{M}_1$ and $\hat{M}_2$, i.e.,

$$(\hat{M}_1, \hat{M}_2) = \phi^{(n)}(\mathbf{Y}). \tag{13}$$

An error occurs whenever $(M_1, M_2) \neq (\hat{M}_1, \hat{M}_2)$.

A rate pair $(R_1, R_2)$ is said to be *achievable* over the Gaussian MAC with conferencing encoders if there exist a sequence of $\{(n, C_{12}, C_{21})\}$-conferences, two sequences of encoding functions $\{\varphi_1^{(n)}, \varphi_2^{(n)}\}$ as in (8) and (9), and a sequence of decoding functions $\{\phi^{(n)}\}$ as in (12) such that the probability of error tends to 0 as the block-length $n$ tends to infinity, i.e.,

$$\lim_{n \to \infty} \Pr\left[(M_1, M_2) \neq (\hat{M}_1, \hat{M}_2)\right] = 0. \tag{14}$$

The *capacity region* $\mathcal{C}$ is defined as the closure of the set of all achievable rate pairs.

We also consider an extension of the setting at hand in the sense of Costa's "Writing on Dirty Paper" channel [2]. Thus, we assume an additional additive Gaussian interference sequence which is non-causally known to both transmitters but not to the receiver. There exist two different scenarios one could envision. A scenario where the transmitters learn the interference sequence *before* the conference, and thus the inputs to the bit-pipes can depend also on the interference; or a scenario where the transmitters learn the interference only *after* the conference. It turns out that the presented results do not depend on which of the two scenarios is considered. The converse we present holds also for the setting where the transmitters know the interference already *before* the conference; and the encoding scheme applies also to the setting where the transmitters know the interference only *after* the conference. In the following we will focus on the setting where the transmitters learn the interference sequences after the conference.

For the setting with interference we need to modify the definitions of the channel in (11) and the encoding functions in (8) and (9); the decoding function remains as in (12).

For given inputs $x_{1,t}$ and $x_{2,t}$ the channel output is given by

$$Y_t = x_{1,t} + x_{2,t} + S_t + Z_t, \tag{15}$$

where the noise sequence $\{Z_t\}$ is defined as before, and where the interference sequence $\{S_t\}$ is an IID zero-mean Gaussian sequence of variance $Q$ and independent of the noise sequence and of the messages.

Denoting the interference sequence by $\mathbf{S} = (S_1, \ldots, S_n)$, the channel input sequences $\mathbf{X}_1$ and $\mathbf{X}_2$ are described as

$$\mathbf{X}_1 = \varphi_{1,\text{IF}}^{(n)}(M_1, \mathbf{V}_2, \mathbf{S}),$$
$$\mathbf{X}_2 = \varphi_{2,\text{IF}}^{(n)}(M_2, \mathbf{V}_1, \mathbf{S}),$$

for some encoding functions $\varphi_{1,\text{IF}}^{(n)}, \varphi_{2,\text{IF}}^{(n)}$ of the form

$$\varphi_{1,\text{IF}}^{(n)}: \mathcal{M}_1 \times \mathcal{V}_{2,1} \times \ldots \times \mathcal{V}_{2,k} \times \mathbb{R}^n \longrightarrow \mathbb{R}^n,$$
$$\varphi_{2,\text{IF}}^{(n)}: \mathcal{M}_2 \times \mathcal{V}_{1,1} \times \ldots \times \mathcal{V}_{1,k} \times \mathbb{R}^n \longrightarrow \mathbb{R}^n.$$

The input sequences are subject to the power constraints (10).

The probability of error, achievable rate pairs, and capacity region $\mathcal{C}_{\text{IF}}$ for this new setting are defined as before.

## II. MAIN RESULTS

*Definition 1:* Define the region

$$\mathcal{C}_\mathcal{G} \triangleq \bigcup_{0 \leq \beta_1, \beta_2 \leq 1} \Bigg\{ (R_1, R_2) :$$
$$R_1 \leq \frac{1}{2}\log\left(1 + \frac{\beta_1 P_1}{\sigma^2}\right) + C_{12},$$
$$R_2 \leq \frac{1}{2}\log\left(1 + \frac{\beta_2 P_2}{\sigma^2}\right) + C_{21},$$
$$R_1 + R_2 \leq \frac{1}{2}\log\left(1 + \frac{\beta_1 P_1 + \beta_2 P_2}{\sigma^2}\right) + C_{12} + C_{21},$$
$$R_1 + R_2 \leq \frac{1}{2}\log\left(1 + \frac{P_1 + P_2 + 2\sqrt{P_1 P_2 \bar{\beta}_1 \bar{\beta}_2}}{\sigma^2}\right) \Bigg\}, \tag{16}$$

where $\bar{\beta}_1 = 1 - \beta_1$ and $\bar{\beta}_2 = 1 - \beta_2$.

*Theorem 1:* The capacity region $\mathcal{C}$ of the Gaussian MAC with conferencing encoders is equal to $\mathcal{C}_\mathcal{G}$,

$$\mathcal{C} = \mathcal{C}_\mathcal{G}. \tag{17}$$

*Remark 1:* The main step in the proof (see Section III-A) is to show that under the Markov condition $X_1\!\!-\!\!\circ\!\!-\!\!U\!\!-\!\!\circ\!\!-\!\!X_2$ the region $\mathcal{R}_{\text{Conf}}(X_1, U, X_2)$ (Definition 2) is maximized by choosing jointly Gaussian distributions on $(X_1, U, X_2)$. A subset of the mutual information expressions which characterize the region $\mathcal{R}_{\text{Conf}}(X_1, U, X_2)$ can be found in the characterization of an achievable region for the MAC with perfect feedback proposed by Cover and Leung [6] and in the characterization of the capacity region of the MAC with common messages derived by Slepian and Wolf [7]. Again, in both characterizations the triple $X_1\!\!-\!\!\circ\!\!-\!\!U\!\!-\!\!\circ\!\!-\!\!X_2$ is required to be Markov, and the same tools as in Section III-A can be used to prove that for Gaussian channels also the Cover-Leung region and the Slepian-Wolf region are maximized by choosing jointly Gaussian distributions on $X_1\!\!-\!\!\circ\!\!-\!\!U\!\!-\!\!\circ\!\!-\!\!X_2$ [4].

*Theorem 2:* The capacity region $\mathcal{C}_{\text{IF}}$ of the Gaussian MAC with conferencing encoders and additive Gaussian interference sequence non-causally known at both encoders equals the capacity region $\mathcal{C}$ of the setting without interference

$$\mathcal{C}_{\text{IF}} = \mathcal{C}. \tag{18}$$

Note that Costa's result [2] on "Writing on Dirty Paper" and Gel'fand and Pinsker's result [8] on "Multi-access Writing on Dirty Paper" are special cases of Theorem 2.

## III. PROOF OF THEOREM 1

The achievability of $\mathcal{C}_\mathcal{G}$, i.e.,

$$\mathcal{C}_\mathcal{G} \subseteq \mathcal{C}, \tag{19}$$

follows by applying the scheme described in [1] with a Gaussian input distribution. The details are omitted.

### A. Converse

To prove the converse, i.e.,

$$\mathcal{C} \subseteq \mathcal{C}_\mathcal{G}, \tag{20}$$

we first outer bound $\mathcal{C}$ by $\mathcal{C}_{\text{Out}}$ (Lemma 1). The converse is then established by showing $\mathcal{C}_{\text{Out}} = \mathcal{C}_\mathcal{G}$. To this end, in Lemma 2 we express the region $\mathcal{C}_\mathcal{G}$ in a similar form to $\mathcal{C}_{\text{Out}}$, i.e., as a union of regions where the union is taken over certain distributions satisfying a Markov condition and power constraints. We then notice that $\mathcal{C}_{\text{Out}}$ and $\mathcal{C}_\mathcal{G}$ differ only with respect to the set of distributions over which the unions are taken (see (21) and (23)): for $\mathcal{C}_{\text{Out}}$ the union is taken over *all* distributions satisfying the Markov condition and the power constraints, and for $\mathcal{C}_\mathcal{G}$ the union is only over those that are *Gaussian*. We conclude the proof by showing that for $\mathcal{C}_{\text{Out}}$ it is sufficient to take the union only over Gaussian distributions (Lemma 3).

*Definition 2:* For a given distribution $p_{X_1UX_2}(\cdot, \cdot, \cdot)$ on the random triple $X_1, U, X_2$, define

$$\mathcal{R}_{\text{Conf}}(X_1, U, X_2)$$
$$\triangleq \Big\{ (R_1, R_2): \quad R_1 \leq I(X_1; Y | X_2 U) + C_{12},$$
$$R_2 \leq I(X_2; Y | X_1 U) + C_{21},$$
$$R_1 + R_2 \leq I(X_1 X_2; Y | U) + C_{12} + C_{21},$$
$$R_1 + R_2 \leq I(X_1 X_2; Y) \quad \Big\},$$

where the mutual informations are computed with respect to the law $p_{UX_1X_2Y}(u, x_1, x_2, y) = p_{X_1UX_2}(x_1, u, x_2)p(y|x_1, x_2)$, where $p(y|x_1, x_2)$ denotes the channel law.

*Definition 3:* Define the region

$$\mathcal{C}_{\text{Out}} \triangleq \bigcup_{\substack{X_1\!-\!\circ\!-\!U\!-\!\circ\!-\!X_2 \\ \mathsf{E}[X_1^2] \leq P_1,\, \mathsf{E}[X_2^2] \leq P_2}} \mathcal{R}_{\text{Conf}}(X_1, U, X_2), \tag{21}$$

where the union is over all joint distributions (not necessarily Gaussian) for which $X_1\!-\!\circ\!-\!U\!-\!\circ\!-\!X_2$ is Markov and for which $\mathsf{E}[X_1^2] \leq P_1$ and $\mathsf{E}[X_2^2] \leq P_2$

*Lemma 1:* The region $\mathcal{C}_{\text{Out}}$ is an outer bound on the capacity region of the Gaussian MAC with conferencing encoders,

$$\mathcal{C} \subseteq \mathcal{C}_{\text{Out}}. \tag{22}$$

*Proof:* Requires only a slight modification of the outer bound in [1] to account for the power constraints. ∎

*Lemma 2:* The region $\mathcal{C}_\mathcal{G}$ in (16) can be expressed as

$$\mathcal{C}_\mathcal{G} = \bigcup_{\substack{X_1^\mathcal{G}\!-\!\circ\!-\!U^\mathcal{G}\!-\!\circ\!-\!X_2^\mathcal{G} \\ \mathsf{E}\left[(X_1^\mathcal{G})^2\right] \leq P_1,\, \mathsf{E}\left[(X_2^\mathcal{G})^2\right] \leq P_2}} \mathcal{R}_{\text{Conf}}(X_1^\mathcal{G}, U^\mathcal{G}, X_2^\mathcal{G}), \tag{23}$$

where the superscript $\mathcal{G}$ is used to indicate that the union is taken only over Gaussian Markov distributions satisfying the second moment constraints.

*Proof:* Follows by evaluating the various mutual information terms in the definition of $\mathcal{R}_{\text{Conf}}$ for Gaussian distributions on $X_1\!-\!\circ\!-\!U\!-\!\circ\!-\!X_2$. ∎

The right hand-sides of (21) and (23) differ only with respect to the set of distributions over which the unions are taken. Therefore, in order to conclude the proof of the converse (20), by Lemma 1 and Equations (21) and (23), it suffices to show that there is no loss in optimality if the union in (21) is taken only over Gaussian Markov triples fulfilling the second moment constraints. This is established by the following Lemma 3.

*Lemma 3:* For any Markov triple $X_1\!-\!\circ\!-\!U\!-\!\circ\!-\!X_2$ fulfilling $\mathsf{E}[X_1^2] \leq P_1$ and $\mathsf{E}[X_2^2] \leq P_2$, there exists a Gaussian Markov triple $X_1^{\mathcal{G}*}\!-\!\circ\!-\!V^{\mathcal{G}*}\!-\!\circ\!-\!X_2^{\mathcal{G}*}$ fulfilling the power constraints $\mathsf{E}\left[(X_1^{\mathcal{G}*})^2\right] \leq P_1$ and $\mathsf{E}\left[(X_2^{\mathcal{G}*})^2\right] \leq P_2$, such that

$$\mathcal{R}_{\text{Conf}}(X_1, U, X_2) \subseteq \mathcal{R}_{\text{Conf}}(X_1^{\mathcal{G}*}, V^{\mathcal{G}*}, X_2^{\mathcal{G}*}). \tag{24}$$

We postpone the proof of Lemma 3 and first state a sequence of definitions and lemmas.

*Lemma 4:* For any (not necessarily Markov) random triple $X_1, U, X_2$ of finite second moments,
$$\mathcal{R}_{\text{Conf}}(X_1, U, X_2) \subseteq \mathcal{R}_{\text{Conf}}(X_1^{\mathcal{G}}, U^{\mathcal{G}}, X_2^{\mathcal{G}})$$
where $(X_1^{\mathcal{G}}, U^{\mathcal{G}}, X_2^{\mathcal{G}})$ is a centered Gaussian vector whose covariance matrix is equal to that of $(X_1, U, X_2)$.

*Proof:* Follows by a conditional version of the Max-Entropy Theorem [3, Theorem 12.1.1.], see also [5]. ∎

*Definition 4:* Define $\mathcal{K}_{\mathcal{G}}$ as the set of $3 \times 3$ positive semi-definite matrices
$$\mathsf{K} = \begin{pmatrix} k_{11} & k_{12} & k_{13} \\ k_{12} & k_{22} & k_{23} \\ k_{13} & k_{23} & k_{33} \end{pmatrix} \quad (25)$$
satisfying one of the two conditions
1) $k_{22} \neq 0$ and $k_{13} k_{22} = k_{12} k_{23}$;
2) $k_{22} = k_{12} = k_{13} = k_{23} = 0$.

*Lemma 5:* A Gaussian triple is Markov if, and only if, its covariance matrix is in $\mathcal{K}_{\mathcal{G}}$.

*Proof:* The "if" direction follows because the law of a Gaussian triple is fully characterized by its mean and its covariance matrix and by noting that for any covariance matrix $\mathsf{K} \in \mathcal{K}_{\mathcal{G}}$ we can construct a Gaussian Markov triple of covariance matrix $\mathsf{K}$. The proof of the later is omitted.

For the proof of the "only if" direction we assume a triple $A$—∘—$B$—∘—$C$ which forms a Markov chain in this order. We distinguish between two cases: $\mathsf{Var}(B) = 0$ and $\mathsf{Var}(B) \neq 0$. If $\mathsf{Var}(B) = 0$, then $B$ is deterministic and the Markov chain $A$—∘—$B$—∘—$C$ implies that $A$ and $C$ are independent. The covariance matrix of $A, B, C$ is then diagonal, and Condition 2) is satisfied. If $\mathsf{Var}(B) \neq 0$, then define $A_0 \triangleq A - \mathsf{E}[A]$, $B_0 \triangleq B - \mathsf{E}[B]$, and $C_0 \triangleq C - \mathsf{E}[C]$ and compute
$$\mathsf{E}[A_0 C_0] = \mathsf{E}[\mathsf{E}[A_0 C_0 | B_0]] = \mathsf{E}[\mathsf{E}[A_0 | B_0] \mathsf{E}[C_0 | B_0]]$$
$$= \mathsf{E}\left[\frac{\mathsf{E}[A_0 B_0]}{\mathsf{Var}(B_0)} B_0 \frac{\mathsf{E}[B_0 C_0]}{\mathsf{Var}(B_0)} B_0\right] = \frac{\mathsf{E}[A_0 B_0] \mathsf{E}[B_0 C_0]}{\mathsf{E}[B_0^2]}. \quad (26)$$
Here, the second equality follows by the Markovity and the third equality by the Gaussianity. By multiplying (26) with $\mathsf{E}[B_0^2] = \mathsf{Var}(B)$ Condition 1) is obtained. ∎

*Lemma 6:* Consider a Markov triple $X_1$—∘—$U$—∘—$X_2$ with $X_1$ and $X_2$ of finite second moments. Let
$$V = \mathsf{E}[X_1 | U] - \mathsf{E}[X_1]. \quad (27)$$
Then, the covariance matrix of the triple $(X_1, V, X_2)$ is in $\mathcal{K}_{\mathcal{G}}$, and
$$\mathcal{R}_{\text{Conf}}(X_1, U, X_2) \subseteq \mathcal{R}_{\text{Conf}}(X_1, V, X_2). \quad (28)$$

*Proof:* The inclusion (28) follows by the following two observations. Exchanging $U$ by a deterministic function of $U$ increases all mutual information expressions in $\mathcal{R}_{\text{Conf}}$ which are conditional on $U$. And changing $U$ does not change the joint distribution of $X_1, X_2$, and hence the unconditional mutual information expression remains the same.

That the covariance matrix of the triple $(X_1, V, X_2)$ is in $\mathcal{K}_{\mathcal{G}}$ follows because (27) and the Markov condition $X_1$—∘—$U$—∘—$X_2$ imply that
$$\mathsf{Cov}[V, X_2] = \mathsf{Cov}[X_1, X_2] \quad \text{and} \quad \mathsf{Cov}[V, X_1] = \mathsf{Var}(V). \quad \blacksquare$$

*Proof of Lemma 3:* Let
$$V \triangleq \mathsf{E}[X_1 | U] - \mathsf{E}[X_1], \quad (29)$$
and define the triple $X_1^{\mathcal{G}*}, V^{\mathcal{G}*}, X_2^{\mathcal{G}*}$ to be zero-mean jointly Gaussian with the same covariance matrix as the triple $X_1, V, X_2$. To conclude the proof we shall show that $X_1^{\mathcal{G}*}$—∘—$V^{\mathcal{G}*}$—∘—$X_2^{\mathcal{G}*}$ forms a Markov chain and that Condition (24) is satisfied. That the triple $X_1^{\mathcal{G}*}$—∘—$V^{\mathcal{G}*}$—∘—$X_2^{\mathcal{G}*}$ forms a Markov chain follows by the Gaussianity of $X_1^{\mathcal{G}*}, V^{\mathcal{G}*}, X_2^{\mathcal{G}*}$ and by Lemma 5, because by Lemma 6 the covariance matrix of $X_1, V, X_2$ is in $\mathcal{K}_{\mathcal{G}}$ and thus, by construction, also the covariance matrix of $X_1^{\mathcal{G}*}, V^{\mathcal{G}*}, X_2^{\mathcal{G}*}$ is in $\mathcal{K}_{\mathcal{G}}$. Note that the triple $X_1, V, X_2$, even though its covariance matrix is in $\mathcal{K}_{\mathcal{G}}$, does not necessarily form a Markov chain in this order, because it is not restricted to be Gaussian.

That Condition (24) is satisfied follows by the following sequence of inclusions:
$$\mathcal{R}_{\text{Conf}}(X_1, U, X_2) \subseteq \mathcal{R}_{\text{Conf}}(X_1, V, X_2)$$
$$\subseteq \mathcal{R}_{\text{Conf}}(X_1^{\mathcal{G}*}, V^{\mathcal{G}*}, X_2^{\mathcal{G}*}), \quad (30)$$
where the first inclusion follows by Lemma 6 and the second inclusion follows by Lemma 4. ∎

## IV. PROOF OF THEOREM 2

The "converse"
$$\mathcal{C}_{\text{IF}} \subseteq \mathcal{C} \quad (31)$$
follows because $\mathcal{C}$ outer bounds the capacity region of the channel with interference even when the interference is also known at the receiver. It remains to prove the "direct part"
$$\mathcal{C} \subseteq \mathcal{C}_{\text{IF}}, \quad (32)$$
i.e., that every rate pair in $\mathcal{C}$ is achievable in the presence of interference. This follows by Lemmas 7 and 8 ahead.

*Definition 5:* Define the region
$$\mathcal{R}_{\text{Ach}} = \bigcup_{0 \leq \beta_1, \beta_2 \leq 1} \Big\{ (R_1, R_2) :$$
$$R_1 \leq \frac{1}{2} \log\left(1 + \frac{\beta_1 P_1}{\sigma^2}\right) + C_{12}, \quad (33)$$
$$R_1 \leq \frac{1}{2} \log\left(1 + \frac{\beta_1 P_1}{\sigma^2}\right)$$
$$+ \frac{1}{2} \log\left(1 + \frac{(\sqrt{\beta_1 P_1} + \sqrt{\bar{\beta}_2 P_2})^2}{\beta_1 P_1 + \beta_2 P_2 + \sigma^2}\right), \quad (34)$$
$$R_2 \leq \frac{1}{2} \log\left(1 + \frac{\beta_2 P_2}{\sigma^2}\right) + C_{21}, \quad (35)$$
$$R_2 \leq \frac{1}{2} \log\left(1 + \frac{\beta_2 P_2}{\sigma^2}\right)$$
$$+ \frac{1}{2} \log\left(1 + \frac{(\sqrt{\bar{\beta}_1 P_1} + \sqrt{\beta_2 P_2})^2}{\beta_1 P_1 + \beta_2 P_2 + \sigma^2}\right), \quad (36)$$
$$R_1 + R_2 \leq \frac{1}{2} \log\left(1 + \frac{\beta_1 P_1 + \beta_2 P_2}{\sigma^2}\right) + C_{12} + C_{21}, \quad (37)$$

$$R_1 + R_2 \leq \frac{1}{2} \log \left( 1 + \frac{P_1 + P_2 + 2\sqrt{P_1 P_2 \bar{\beta}_1 \bar{\beta}_2}}{\sigma^2} \right) \right\}, \quad (38)$$

where $\bar{\beta}_1 = 1 - \beta_1$ and $\bar{\beta}_2 = 1 - \beta_2$.

*Lemma 7:* The capacity region $\mathcal{C}_{\text{IF}}$ includes $\mathcal{R}_{\text{Ach}}$:

$$\mathcal{C}_{\text{IF}} \supseteq \mathcal{R}_{\text{Ach}}. \quad (39)$$

*Proof:* See Section IV-A. ∎

*Lemma 8:* The achievable region $\mathcal{R}_{\text{Ach}}$ equals the capacity region $\mathcal{C}$ of the Gaussian MAC with conferencing encoders,

$$\mathcal{R}_{\text{Ach}} = \mathcal{C}.$$

### A. Coding Technique Achieving $\mathcal{R}_{Ach}$

In this section we sketch a coding technique that achieves the region $\mathcal{R}_{\text{Ach}}$. The analysis is omitted.

The two transmitters first create a common message by communicating over the pipes as in [1]. Thus, after the conference Transmitter 1 is cognizant of the common message and of an independent private message. It allocates power $(1 - \beta_1)P_1$ to the common message and power $\beta_1 P_1$ to the private message. Similarly for Transmitter 2.

The coding technique involves time-sharing between two schemes. Both schemes apply successive decoding at the receiver, where the receiver first decodes the common message followed by the private messages. But they differ in the decoding order of the private messages. We describe the scheme where the decoding of the common message is followed by the decoding of the private message of Transmitter 1 and only thereafter by the decoding of the private message of Transmitter 2. The other scheme where the decoding of the common message is followed by the private message of Transmitter 2 is analogous.

We first describe the encoding of the common message. Before transmission begins, the transmitters agree on a (single-user) dirty-paper code for power $P_0 \triangleq \left(\sqrt{(1-\beta_1)P_1} + \sqrt{(1-\beta_2)P_2}\right)^2$, noise variance $(\beta_1 P_1 + \beta_2 P_2 + \sigma^2)$, and interference $\mathbf{S}$. Transmitter 1 encodes the common message using this dirty-paper code and scales the resulting sequence by $\frac{\sqrt{(1-\beta_1)P_1}}{\sqrt{P_0}}$. Transmitter 2 encodes the common message with the same code, but scales the resulting sequence by $\frac{\sqrt{(1-\beta_2)P_2}}{\sqrt{P_0}}$. (The channel will coherently combine the two sequences.)

Independently of the common message, the transmitters encode the private messages. Transmitter 1 encodes its private message using a dirty-paper code for power $\beta_1 P_1$, noise variance $\sigma^2 + \beta_2 P_2$, and interference $\mathbf{S}_1 \triangleq \left(1 - \frac{\left(\sqrt{(1-\beta_1)P_1} + \sqrt{(1-\beta_2)P_2}\right)^2}{P_1 + P_2 + 2\sqrt{(1-\beta_1)(1-\beta_2)P_1 P_2} + \sigma^2}\right) \mathbf{S}$. Transmitter 2 encodes its private message with a dirty-paper code for power $\beta_2 P_2$, noise variance $\sigma^2$, and interference $\mathbf{S}_2 \triangleq \left(1 - \frac{\beta_1 P_1}{\beta_1 P_1 + \beta_2 P_2 + \sigma^2}\right) \mathbf{S}_1$. Each transmitter sends the sum of the two sequences produced for the common message and for its private message.

The receiver performs successive decoding. It starts by decoding the common message based on nearest neighbor decoding as in [9], while treating the sequences which the transmitters produced for the private messages as additional noise. Then, the receiver subtracts (or "strips off") the decoded common-message codeword from the channel outputs and proceeds to decode the private message of Transmitter 1. (Here, the "common-message codeword" is not the resulting sequence of the dirty-paper code, but the codeword in the bin of the common message which was selected during the encoding procedure of the dirty-paper code.) To decode the private message of Transmitter 1, the receiver again uses nearest neighbor decoding and treats the sequence which Transmitter 2 produced for its private message as additional noise. Finally, it subtracts the decoded Transmitter 1-private-message codeword and decodes the private message of Transmitter 2.

*Remark 2:* To encode the different messages the transmitters use dirty-paper codes for scaled versions of the interference $\mathbf{S}$. The reason for this is that the codewords that the receiver subtracts depend on the interference sequence $\mathbf{S}$, and the resulting channel seen in subsequent decoding phases is interfered by a scaled version of $\mathbf{S}$.

*Remark 3:* Our coding scheme is different from Willems's scheme [1] (for the setting without interference). In his scheme the transmitters apply superposition coding and the receiver applies joint decoding. Our approach has two advantages: It simplifies the analysis, and it achieves the same result also if the noise sequence $\{Z_t\}$ is not IID Gaussian but any arbitrary ergodic process of second moment $\sigma^2$. However, our approach leads to the additional constraints (34) and (36). Fortunately, by Lemma 8, these additional constraints don't shrink the resulting region.

### ACKNOWLEDGMENT

We would like to thank Stephan Tinguely, İ. Emre Telatar, and Tsachy Weissman for helpful discussions.